\documentclass[12pt,preprint]{aastex}

\usepackage[bf,small]{caption}

\newcommand{\microns}{$\mu$m}

\def\deg{$^{\circ}$}
\def\ga{\mathrel{\hbox{\rlap{\hbox{\lower4pt\hbox{$\sim$}}}\hbox{$>$}}}}
\def\la{\mathrel{\hbox{\rlap{\hbox{\lower4pt\hbox{$\sim$}}}\hbox{$<$}}}}
\def\msunyr{$M$ \mbox{$_{\normalsize\odot}$}\rm{yr}$^{-1}$}
\def\msun{$M$\mbox{$_{\normalsize\odot}$}}

\def\kms{\,km~s$^{-1}$}

\def\arcsec{$^{\prime \prime}$}

\slugcomment{}

\shorttitle{Chemical abundances of Galactic Centre RSGs}
\shortauthors{Davies et al.}

\begin{document}


\title{The chemical abundances in the Galactic Centre from the
  atmospheres of Red Supergiants}


\author{Ben Davies\altaffilmark{1,2}, Livia Origlia\altaffilmark{3},
Rolf-Peter Kudritzki\altaffilmark{4}, Don F.\ Figer\altaffilmark{2}, 
R.\ Michael Rich\altaffilmark{5}, Francisco Najarro\altaffilmark{6} }

\affil{$^{1}$School of Physics \& Astronomy, University of Leeds,
  Woodhouse Lane, Leeds LS2 9JT, U.K.}
\affil{$^{2}$Chester F.\ Carlson Center for Imaging Science, Rochester
Institute of Technology, 54 Lomb Memorial Drive, Rochester NY, 14623,
USA} 
\affil{$^{3}$INAF-Osservatorio Astronomico di Bologna, Via Ranzani 1, I-40127 Bologna, Italy}
\affil{$^{4}$Institute for Astronomy, University of Hawaii, 2680
Woodlawn Drive, Honolulu, HI, 96822, USA} 
\affil{$^{5}$Department of Physics and Astronomy, University of California at Los Angeles, Los Angeles, CA 90095-1547, USA}
\affil{$^{6}$Instituto de Estructura de la Materia, Consejo Superior
  de Investigaciones Cientificas, Calle Serrano 121, 28006 Madrid,
  Spain.}

\begin{abstract}
The Galactic Centre (GC) has experienced a high degree of recent
star-forming activity, as evidenced by the large number of massive
stars currently residing there. The relative abundances of chemical
elements in the GC may provide insights into the origins of this
activity. Here, we present high-resolution $H$-band spectra of two Red
Supergiants in the GC (IRS~7 and VR~5-7), and in combination with
spectral synthesis we derive abundances for Fe and C, as well as other
$\alpha$-elements Ca, Si, Mg Ti and O. We find that the C-depletion in
VR~5-7 is consistent with the predictions of evolutionary models of
RSGs, while the heavy depletion of C and O in IRS~7's atmosphere is
indicative of deep mixing, possibly due to fast initial rotation
and/or enhanced mass-loss. Our results indicate that the {\it current}
surface Fe/H content of each star is slightly above Solar. However,
comparisons to evolutionary models indicate that the {\it initial}
Fe/H ratio was likely closer to Solar, and has been driven higher by
H-depletion at the stars' surface.  Overall, we find $\alpha$/Fe
ratios for both stars which are consistent with the thin Galactic
disk. These results are consistent with other chemical studies of the
GC, given the precision to which abundances can currently be
determined. We argue that the GC abundances are consistent with a
scenario in which the recent star-forming activity in the GC was
fuelled by either material travelling down the Bar from the inner
disk, or from the winds of stars in the inner Bulge -- with no need to
invoke top-heavy stellar Initial Mass Functions to explain anomalous
abundance ratios.

\end{abstract}



\section{Introduction} \label{sec:intro}
Understanding the star-formation history at the Galactic Centre (GC)
is key to understanding the Galaxy's secular evolution. The inner
40\,pc has clearly experienced a high degree of recent star formation,
apparent from the young stellar clusters residing there.  These
clusters host some of the most massive stars known to exist in the
Galaxy -- the Arches cluster \citep{Figer02}, the Quintuplet cluster
\citep{Figer99} and the Central cluster \citep{Paumard06}. It remains
an open question how these clusters were formed, given the harsh
environment of the GC, or whether they were formed {\it in situ} in
the GC at all.

The key to answering this question may lie in chemical abundance
measurements. By probing the relative abundances of certain elements
we may infer the star formation history of the Galactic
Centre. Broadly, $\alpha$-elements (such as Mg, Ti, Ca, Si) are
produced predominantly in the core-collapse supernovae (CCSNe) of
massive stars. Meanwhile, Fe-peak elements are produced in Type-Ia
SNe, that is the thermonuclear explosion of a low-mass white dwarf
which has accreted a sufficient amount of mass from a binary companion
to reach the Chandrasekhar limit. Hence, $\alpha$ enrichment happens
on much shorter timescales ($\sim$10 Myrs), due to the comparitively
brief lifetimes of massive stars; while Fe-enrichment occurs on much
longer timescales ($\sim$Gyrs). In the Galaxy, the ratio of
$\alpha$/Fe is found to be lower in the thin disk than in the other
Galactic environments of the bulge and thick disk
\citep[e.g.][]{Bensby04,R-O05,Luck06}. This is commonly explained as
being due to the rapid formation of the bulge and halo, where star
formation in the early Galaxy was intense but brief, and the elevated
$\alpha$ abundances from the early CCSNe were `frozen-in'. In the thin
disk however, star formation has continued throughout the lifetime of
the Galaxy, allowing Type-Ia SNe to contribute to the chemical
evolution, and driving the $\alpha$/Fe ratio to below that of the halo
and bulge. Meanwhile, a gradient of increasing metal content is found
at lower Galacto-centric distances ($R_{\rm GC}$) within the disk
\citep[e.g.][]{Luck06}, suggesting that the rate of star formation in
the thin disk has been higher toward the GC. However, the absence of
suitable probes of metallicity inwards of $R_{\rm GC} \approx 4$kpc,
combined with the large extiction towards the GC, have meant that it
has been unclear whether the trend of increasing metal content flows
all the way to the center of the Galaxy.

In recent years, advancements in infra-red astronomy have meant that
it is now possible to make abundance measurements of stars in the GC,
in order to place this region within the evolutionary framework of the
whole Galaxy. Infra-red studies are necessary to work around the 30
magnitudes of visual extinction toward the GC, with many such studies
concentrating on the large number of high-mass stars present
there. Massive stars provide excellent probes of chemical abundances;
their spectra are rich in emission/absorption lines, while their short
lifetimes mean that they provide an up-to-date picture of the local
metallicity.

In studies of hot stars, \citet{Najarro04} and \citet{Martins08} both
attempted to infer the initial O abundance of the Arches cluster from
the asymptotic nitrogen abundances of the cluster's most evolved
stars. When comparing the results of both studies to the latest
stellar evolutionary models, in which the relative abundance ratios of
the heavy elements are fixed, each study suggests a slightly
super-Solar global metal content $Z/Z_{\odot}$, though neither study
was able to give abundances of specific $\alpha$ or Fe-peak
elements. In a study of two Luminous Blue Variables (LBVs) in the
Quintuplet cluster \citet{Najarro08} was able to make direct
measurements of Fe content, plus the $\alpha$-elements Si and Mg. They
found Fe to be approximately Solar, while Si and Mg were enhanced by
$\approx$ 0.3$\pm$0.2dex with respect to Solar
values. \citet{Geballe06} studied the star IRS~8 in the Central
cluster using low-resolution $K$-band spectroscopy. They argued that
the 2.116\microns\ feature, which is thought to be a blend of C~{\sc
iii}, N~{\sc iii} and O~{\sc iii} transitions, was sensitive to O
content, and used it to derive an O abundance which was super-Solar at
the 2$\sigma$ level when compared to the \citet{Asplund05} Solar
abundances\footnote{When the abundances from the 1-D Solar model of
\citet{G-S98} are used, the derived O abundance is consistent with
Solar.}.

There have also been several studies of cool massive stars,
specifically Red Supergiants (RSGs). \citet{Carr00} performed the
first abundance study of a star in the GC, modelling the high
resolution $H$ and $K$ spectrum of the Red Supergiant (RSG) in the
Central cluster IRS~7. They found approximately Solar values for Fe to
within their quoted uncertainties, with a depletion of O with respect
to solar likely due to internal CNO processing. \citet{Ramirez00}
analysed a further five RSGs in the central cluster, using $K$-band
high resolution data, again finding Fe content consistent with
Solar. Ramirez et al.\ also studied the Quintuplet RSG VR5-7, finding
an Fe abundance consistent with the rest of their
sample. \citet{Cunha07} analysed this same set of stars as the Ramirez
et al.\ sample, adding $H$-band spectra to the dataset. Again, Solar
Fe content was found, but they also found marginal evidence for
$\alpha$ enhancement: the ratios of both O/Fe and Ca/Fe were found to
be above Solar at the 2$\sigma$ level, again in comparison to the
Asplund et al.\ Solar values.

Studies of GC H{\sc ii} regions have also yielded marginal evidence
for super-Solar $\alpha$ abundances
\citep{Simpson95,Rudolph06}. Analysis of X-ray spectroscopy of the
Sg~A~East supernova (SN) remnant by \citet{Sakano04} shows strong
evidence for $\alpha$-enrichment, however if the object is a remnant
of a core-collapse SN one expects to see significant self-enrichment
of the $\alpha$-elements.

The results described above are all somewhat consistent with a picture
of chemical abundances in the GC which are Solar to within the
uncertainties, with marginal evidence for
$\alpha$-enhancement. Several authors have suggested that the
super-Solar $\alpha$/Fe ratio is real, and is indicative of the star
forming history in the GC. Several explanations have been suggested to
explain this result:

\begin{itemize}
\item A recent burst of star-formation produced a great number of
  massive stars, which evolved to core-collapse SNe. This has enriched
  the local environment in $\alpha$-elements, an effect which can now
  be seen in the surface abundances of subsequent generations of
  stars. However, this is generally inconsistent with the mix of old,
  intermediate and young stellar populations found in the GC. Analyses
  of the GC's H-R diagram and luminosity function have led various
  authors to argue for a constant rate of star-formation over the last
  $\sim$Gyr, rather than an isolated burst
  \citep[e.g.][]{Blum03,Figer04,Maness07}.

\item The harsh environment of the GC results in a Jeans mass larger
  than in the rest of the Galaxy, and produces a top-heavy Initial
  Mass Function (IMF) \citep{Morris93}. The increased rate of CCSNe
  due to the higher fraction of massive stars results in a $\alpha$/Fe
  ratio above that of the disk. This is consistent with the results of
  \citet{Maness07}, who argued for a top-heavy IMF when comparing the
  observed GC H-R diagram with quantitative model predictions.

\item The GC's source of star-forming material is provided by the
  winds of Red Giants in the Bulge, which tend to have high
  $\alpha$/Fe ratios. Thus the natal material was
  already $\alpha$-enriched before it arrived at the GC \citep{M-S96}.
\end{itemize}

However, as stated above, the evidence for a high $\alpha$/Fe ratio is
marginal. Works which study both the Fe and $\alpha$ content directly
are limited to \citet{Cunha07} and \citet{Najarro08}; the former
studied O and Ca, while the latter studied Si and Mg. Oxygen can be a
poor tracer of $\alpha$ abundances in massive stars, as to some extent
it is processed by nuclear burning and can give results which are not
representative of the other $\alpha$ elements.

Here we present an abundance study of two of the GC RSGs, IRS~7 in the
Central cluster and VR~5-7 in the Quintuplet cluster. For the first
time we study a range of $\alpha$-elements -- Ca, Si, Mg, O and Ti --
while also studying Fe and C.

Our method is both independent and complementary to the previous
studies of these stars by \citet{Carr00}, \citet{Ramirez00} and
\citet{Cunha07}, providing full spectral synthesis of the observed
spectra and equivalent width measurements of selected lines.  This
method has been proven to be effective in determining chemical
abundances of low mass giants in the Galactic bulge
\citep[e.g.][]{Origlia02,Origlia05a,Rich07}
as well as of young stellar clusters dominated by red supergiants
\citep{Larsen06,Larsen08}.

We begin in Sect.\ \ref{sec:obs} with a description of our
observations and data-reduction steps, followed by a description of
our abundance analysis methods in Sect.\ \ref{sec:an}. The results are
presented and discussed in Sect.\ \ref{sec:results}.

\section{Observations \& data reduction} \label{sec:obs}
Observations were taken with NIRSPEC, the cross-dispersed echelle
spectrograph mounted on Keck-II, during the night of 3rd June 1999. We
used the instrument in high-resolution mode, with the NIRSPEC-5
filter, in conjunction with the 0.576\arcsec\ $\times$ 12\arcsec\
slit. The dispersion angle was set to 62.53\deg, with cross-dispersion
angle set to 35.53\deg. This gave us a spectral resolution of
$\sim$17,000 of select regions in the wavelength range of
1.5-1.7\microns.

We integrated on each star for 20s in each of two nod-positions along
the slit. In addition to the cluster stars, we also observed
HD~104337, a B2~IV star, as a telluric standard on each
night. Flat-fields were taken with a continuum lamp. For wavelength
calibration purposes, arc frames were taken with Ar, Ne, Xe and Kr
lamps to provide as many template lines as possible in the narrow
wavelength range of each spectral order. As high-precision radial
velocity measurements were not required for this study, no etalon
frames were taken.

We subtracted nod-pairs of frames to remove the sky background, dark
current and detector bias level. Each frame was then flat-fielded
using the continuum-lamp exposures. We corrected for the warping of
each order in a process known as {\it rectification}. The warping in
the spatial and dispersion directions were characterized by fitting
3rd-degree polynomials to the two star-traces in a nod-pair and the
arc lines respectively. These fits were then used to resample each
order onto a linear grid. As the arc-line wavelengths are known, this
process also wavelength-calibrates the data.

Cosmic ray hits and bad pixels were identified by taking the ratio of
two spectra of the same object, and identifying values lying outside
5$\sigma$ of the residual spectrum. These values were replaced with
the median value of the three neighbouring pixels either side. We
removed the H and He~{\sc i} absorption features of the telluric
standard via linear interpolation. The atmospheric absorption features
in the science frames were then removed by dividing through by the
telluric standard. Finally, the spectra were normalised by dividing
through by the median continuum value. The dense absorption-line
spectra make signal-to-noise ratio (SNR) estimates difficult, but we
estimate from weak features we know are real that the SNR is better
than 100 for all stars observed.

\subsection{Data analysis} \label{sec:an}
Abundance analysis is performed by using full spectral synthesis
techniques and equivalent width measurements of representative lines.
Indeed, at the NIRSPEC resolution of ~17,000 a few single
roto-vibrational OH lines and CO bandheads can be measured and used to
derive accurate oxygen and carbon abundances.  Abundances of other
metals can be derived from the atomic lines of Fe~I, Mg~I, Si~I, Ti~I,
and Ca~I.

Photometric estimates of the stellar parameters are initially used as
input to compute a grid of synthetic spectra of red supergiant stars
with varying atmospheric parameters and abundances, by using an
updated \citep{Origlia02,Origlia03} version of the code described in
\citet{Origlia93}.  Briefly, the code uses the LTE approximation and is
based on molecular blanketed model atmospheres of \citet{Johnson80}.  The
code includes several thousands of near IR atomic lines and molecular
roto-vibrational transitions due to CO, OH and CN.  Three main
compilations of atomic oscillator strengths are used, namely the
Kurucz database\footnote{\tt
http://cfa-www.harward.edu/amdata/ampdata/kurucz23/sekur.html}, and
those published by \citet{B-G73} and \citet{M-B99}.  

Abundance estimates are mainly obtained by best-fitting the full
observed spectrum and by measuring the equivalent widths of a few
selected features, dominated by a specific chemical element -- see
Table \ref{tab:ew} for a list of some of the the diagnostic lines,
along with their oscillator stregths and excitation potentials. The
H-band spectra of RSGs contain thousands of absorption lines, and we
concentrate on fitting those spectral features which are relatively
unblended and which have reliable atomic/molecular data. We do not
attempt to fit features which are blends of poorly-understood
transitions. Typical equivalent width values of fitted features range
between 300 and 800 m\AA\ with a conservative error of
$\pm$20~m\AA\ to also account for a $\pm$2\% uncertainty in the
continuum positioning.

For initial estimates of $T_{\rm eff}$ we use the values quoted by
\citet{Cunha07}. We also use the intrinsic luminosities determined in
this paper in combination with the Geneva evolutionary models to
estimate the surface gravity $\log g$. The values of $\log g$ and the
microturbulent velocity $\xi$ are fine-tuned using the CO and OH
lines, which are sensitive to these parameters. We found that $\log g
= 0.0$ and $\xi = 3$\kms\ provided good fits for both stars.

We note that the parameter of macroturbulence is not specifically
included in our current analysis. In previous spectral synthesis
studies of the objects in this paper authors have required the
inclusion of macroturbulent velocities in the region of
15-25\kms. However, as these values are comparable to the spectral
resolution of our observations (18\kms), we do not attempt to
constrain this parameter. We do find some evidence for an additional
broadening component in the spectrum of IRS~7, this is discussed
further in Sect.\ \ref{sec:results}.

\begin{figure}[t]
  \centering
  \includegraphics[width=12cm]{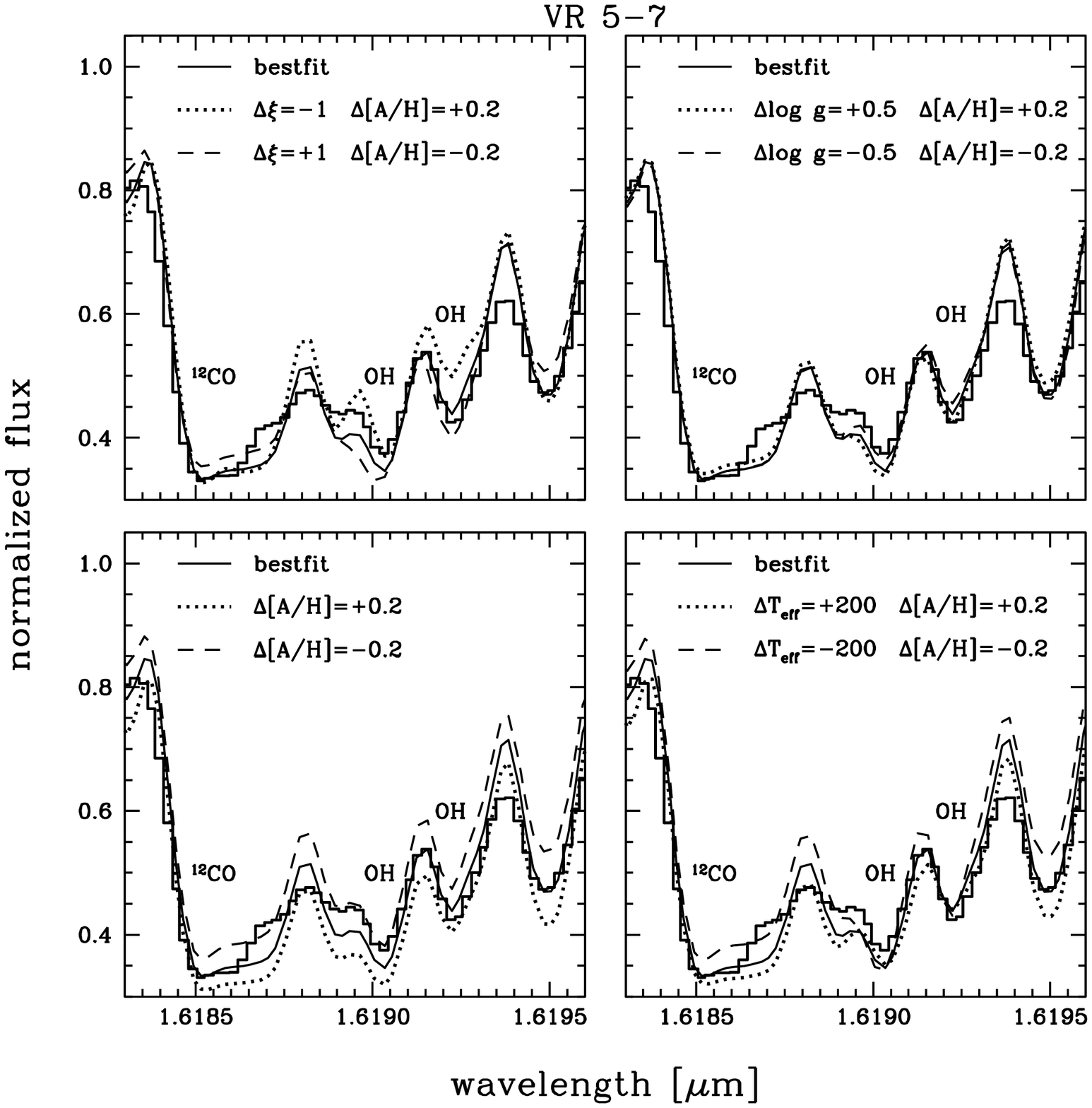}
  \caption{The observed spectrum of VR~5-7 ({\it histogram})
    overplotted with the best-fit model ({\it solid-line}) and test
    models, which vary the stellar parameters and abundances,
    illustrating the robustness of the best-fit solution. Bottom-left
    panel: varying abundances only; bottom right panel: effective
    temperature varied and abundances adjusted accordingly; top-left
    panel: microturbulence varied with abundances adjusted; top-right
    panel: gravity varied and abundances adjusted.}
  \label{fig:err}
\end{figure}

The model which better reproduces the overall observed spectrum and
the equivalent widths of selected lines is chosen as the best fit
model for that particular spectrum. In order to check the statistical
significance of our best-fit solution, we compute a set of six {\it
  test} models with varying stellar parameters of T$_{\rm
  eff}$=$\pm$200\,K, $\log g = \pm$0.5dex, and $\xi = \pm$1\kms. For
each test model, the abundances $A(X)$ are fine-tuned to produce the
best fit in that model, typically by $\pm$0.2dex. We then compute the
residuals between these models and the observed spectrum. We find that
our best-fit model always produces significantly lower residuals than
the test models (see Fig.\ \ref{fig:err}). This process was repeated
with smaller variations in the stellar parameters of the test models
until we found a set of test models which produced residuals
comparable to those of the best-fit model. We found that models with
$\delta$T$_{\rm eff}$=$\pm$100\,K, $\delta (\log g) = \pm$0.3dex) and
$\delta \xi = \pm$0.5\kms\ had the same statistical significance as
our best-fit model . We adopt these values as the experimental
uncertainties in each parameter. The uncertainties in abundance levels
varies from element to element, and depends on other factors such as
blending and continuum placing. The abundance uncertainties are
typically in the range 0.1-0.2dex.

\begin{figure}[p]
  \centering
  \includegraphics[width=15cm]{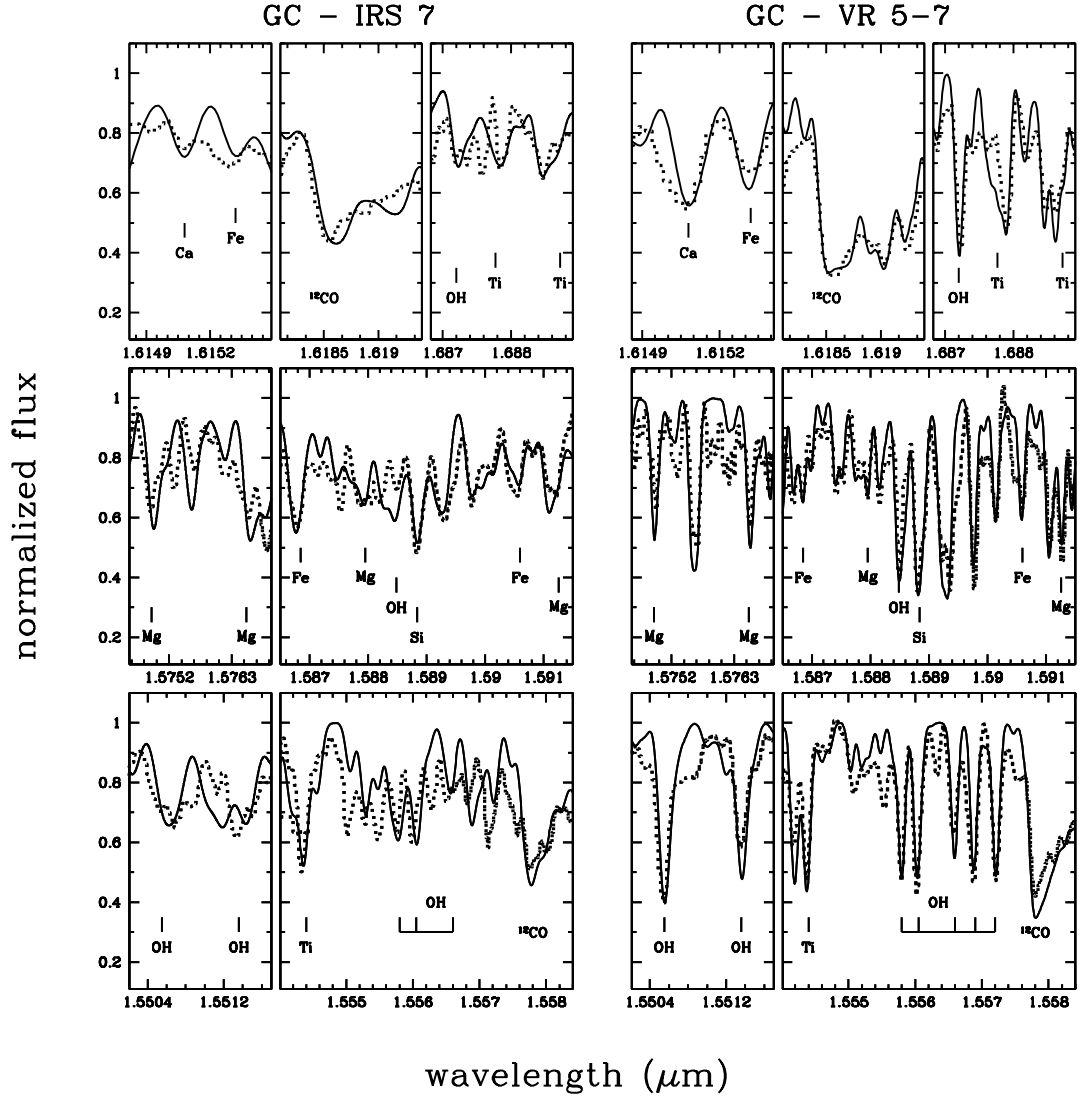}
  \caption{Select regions of the two stars studied (dotted-line),
    overplotted with the best-fitting model for each (solid-line). The
    left panel shows IRS~7 in the Central Cluster; the right panel
    shows VR~5-7 in the Quintuplet Cluster.}
  \label{fig:spec}
\end{figure}


\begin{table*}
\scriptsize
\begin{center}
\caption[]{Oscillator strengths (Log {\it gf}), excitation potentials
  ($\chi$ in eV) and equivalent widths (EW) of some unblended
  representative lines for the observed stars in the Galactic Centre.}
\label{tab:ew}
\begin{tabular}{lcccccccc}
\hline\hline
 & Ca~{\sc i} & Fe~{\sc i} & Fe~{\sc i} & Mg~{\sc i} & Si~{\sc i} & OH & OH & Ti~{\sc i} \\
 & $\lambda $1.61508& $\lambda $1.61532&$\lambda $1.55317&$\lambda $1.57658&$\lambda $1.58884&$\lambda $1.55580&$\lambda $1.55036&$\lambda $1.55437\\
\hline \\
Log {\it gf} &0.362&-0.821&-0.357&0.380&-0.030&-5.492&-7.687&-1.480\\ 
$\chi$ (eV) &5.302&5.35 &5.64&5.93&5.08&0.30&0.84&1.88\\ \\
{\it EW (m\AA)} \vspace{2mm} \\
~~IRS7 & 412 & 420 & 360 & 520 & 745 & 529 & 505 & 352 \\
~~VR5-7& 468 & 410 & 303 & 532 & 704 & 597 & 628 & 345 \vspace{2mm}  \\
\hline
\end{tabular}
\end{center}
\end{table*}



\begin{deluxetable}{lcccccccccccccc}
\tabletypesize{\scriptsize}
\tablecaption{Derived temperatures and abundances for the stars studied, and
 comparisons to previous works. \label{tab:results}}
\tablewidth{0pt}
\tablehead{
\colhead{Reference} &
\colhead{$\rm T_{eff} (K)$}&
\colhead{$\log g$}&
\colhead{$\xi$/\kms}&
\colhead{A(Fe)}&
\colhead{A(O)}&
\colhead{A(Si)}&
\colhead{A(Mg)}&
\colhead{A(Ca)}&
\colhead{A(Ti)}&
\colhead{A(C)}
}
\startdata
\multicolumn{9}{l}{\small IRS 7} \vspace{1mm}  \\
          This work &   3600   &  0.0 & 3.0 & 7.59   &  8.11   &  7.75   &  7.53   &  6.46   &  5.22   & 7.77 \\
\vspace{1mm}     &  $\pm$ 200  & $\pm$ 0.3 & $\pm$ 0.5 & $\pm$ 0.10   & $\pm$ 0.13   & $\pm$ 0.18   & $\pm$ 0.15   & $\pm$ 0.16   & $\pm$ 0.11   & $\pm$ 0.07 \\

   Carr et al.\ 2000 &   3600   & -0.6 & 3.0 & 7.50   & 8.13   &  --   & --   &  --   &  --   & 7.78 \\
\vspace{1mm}  & $\pm$ 230 & $\pm$ 0.2  & $\pm$ 0.3 & $\pm$ 0.13   & $\pm$ 0.32   & --   & --   & --   & --   & $\pm$ 0.13 \\

Cunha et al.\ 2007  &   3650   & -0.5 & 3.2  & 7.66   &  --     &  --     &  --     &  6.86   &  --     & --   \\
\vspace{1mm} & $\pm$ 150  & $\pm$ 0.3 & $\pm$0.3 & $\pm$ 0.15   & --   & --   & --   & $\pm$ 0.15   & --   & -- \\     

Ramirez et al.\ 2000  &   3470   & -0.6 & 3.3 &  7.61   &  --     &  --     &  --     &  --   &  --     & --  \\
\vspace{3mm}   &  $\pm$ 250   & $\pm$ 0.2 & $\pm$0.4 & $\pm$ 0.27   & --   & --   & --   & --   & --   & -- \\     

\multicolumn{9}{l}{\it Average Central Cluster abundances:} \vspace{1mm}  \\
Cunha et al.\ 2007 & -- & -- &  -- & 7.59  &  9.04  &  --  &  --  &  6.57  &  --  &  -- \\ 
\vspace{1mm}   &  --  & -- & -- & $\pm$ 0.15  & $\pm$ 0.19   & --   & --   & $\pm$ 0.14  & --   & -- \\
Ramirez et al.\ 2000 & -- & -- & -- & 7.61  &  --  &  --  &  --  &  --  &  --  &  -- \\ 
\vspace{1mm}   &  --  & -- & -- & $\pm$ 0.22  & --   & --   & --   & --  & --   & -- \\
\\
\hline
\\
\multicolumn{9}{l}{\small VR\,5-7} \vspace{1mm}  \\
   This work &   3400   &  0.0 & 3.0 & 7.55   & 9.09   &  7.65   & 7.53   &  6.50   &  5.15   & 8.22 \\
\vspace{1mm}    &  $\pm$ 200  & $\pm$ 0.3 & $\pm$ 0.5 & $\pm$ 0.10   & $\pm$ 0.11   & $\pm$ 0.14   & $\pm$ 0.14   & $\pm$ 0.13   & $\pm$ 0.09   & $\pm$ 0.07 \\\

Cunha et al.\ 2007  &   3600 & -0.15 & 2.6  &  7.60   &  --     &  --     &  --     &  6.51   &  --     & --   \\
\vspace{1mm}          &  $\pm$ 150  & $\pm$ 0.3 & $\pm$ 0.3 & $\pm$ 0.15   & --   & --   & --   & $\pm$ 0.15   & --   & -- \\     

Ramirez et al.\ 2000  &   3500 & -0.2 & 2.9  &  7.61   &  --     &  --     &  --     &  --   &  --     & --   \\
\vspace{3mm}  &  $\pm$ 300  & $\pm$0.3 & $\pm$0.5 & $\pm$ 0.23   & --   & --   & --   & --   & --   & -- \\     

\multicolumn{9}{l}{\it Average Quintuplet Cluster abundances:} \vspace{1mm}  \\
Najarro et al.\ 2008 & -- & -- & -- &  7.54  &  --  &  7.85  &  7.84  &  --  &  --  &  --  \\
\vspace{1mm} &  --  & -- & -- & $\pm$ 0.15   & --   & $\pm$ 0.25  & $\pm$ 0.20   &  --   & --   & -- \\
\\
\hline
\\
\multicolumn{9}{l}{\small Solar values} \vspace{1mm}  \\
Grevesse \& Sauval 1998 & -- & -- & -- &  7.50  &  8.83  &  7.55  &  7.52  &
6.36  &  5.02  &  8.57  \\
    \vspace{1mm}     & -- & -- & -- & $\pm$ 0.05 & $\pm$ 0.06 & $\pm$ 0.05 &
$\pm$ 0.05 & $\pm$ 0.02 & $\pm$ 0.06 & $\pm$ 0.06 \\

Asplund et al.\ 2005 & -- & -- & -- &  7.45  &  8.66  &  7.51  &  7.53  &  6.31
&  4.90  &  8.39  \\
    \vspace{1mm}     & -- & -- & -- & $\pm$ 0.05 & $\pm$ 0.05 & $\pm$ 0.04 &
$\pm$ 0.09 & $\pm$ 0.04 & $\pm$ 0.06 & $\pm$ 0.05 \\
\enddata
\end{deluxetable}

\section{Results \& Discussion} \label{sec:results}
In Fig.\ \ref{fig:spec} we show select regions of each stars' spectrum
containing diagnostic lines, with the best-fit model overplotted. In
the case of VR~5-7 we found we were able to fit the diagnostic lines
without the inclusion of any additional broadening. Therefore, any
macroturbulence present must be below or comparable to the
instrumental broadening -- the spectral resolution of our
observations, 18\kms. This is consistent with previous studies of this
object at higher spectral resolution, which have required
macroturbulent velocities of 12-15\kms\ \citep{Ramirez00,Cunha07}.

For IRS~7, we found that the fit could be improved by convolving the
spectrum with a broadening profile -- either using a Doppler profile
of 20\kms, or by increasing the gaussian (instrumental) broadening to
27\kms. These values are consistent with macroturbulent velocities of
20-25\kms\ used by Ramirez et al.\ and Cunha et al.\ in their fits to
higher resolution spectra. The fit to IRS~7 shown in
Fig.\ \ref{fig:spec} includes this extra Doppler broadening. In
general, we found that we were unable to achieve the same quality of
fit for IRS~7 as for VR~5-7, and this is reflected in the slightly
increased uncertainties in the derived elemental abundances. One
possible explanation for this could be the interaction of the star's
outer atmosphere with nearby hot stars, augmenting the ionization
structure and/or inducing departures from spherical symmetry in
IRS~7's envelope \citep{Serabyn91,Y-Z91,Y-Z92}.

In Table \ref{tab:results}, the derived abundances are shown for each
element studied, as well as the derived stellar temperature. We find
temperatures gravities and microturbulent velocities for each star
which are in good agreement with previous studies of the same
objects. We quantify the abundance of each element $X$ in the form
$A({X}) = \log(X/{\rm H}) + 12$. We also tabulate the abundances
derived for the same stars by other authors using independent
methods. Finally, we also show the average abundances derived for the
stars' host clusters from other independent studies in the
literature. Below, we describe the elemental abundances individually.

\subsection{Iron}
From Table \ref{tab:results} we see that the Fe abundances we derive
are in excellent agreement with previous studies of the same
objects. They are also consistent with the average abundances of their
host clusters: our measurement of $A({\rm Fe}) = 7.59 \pm 0.10$ for
IRS~7 agrees perfectly with the average $A$(Fe) of 10 RSGs in the
Central cluster studied by \citet[][$\bar{A}({\rm
Fe})=7.59\pm0.15$]{Cunha07}, and the 9 RSGs observed by
\citet[][$\bar{A}({\rm Fe})=7.61\pm0.22$]{Ramirez00}. Our measurement
for VR~5-7 ($A({\rm Fe}) = 7.55 \pm 0.10$) is also consistent with the
average of the two LBVs studied by \citet[][$\bar{A}({\rm Fe}) = 7.54
\pm 0.15$]{Najarro08}.

When comparing the measured Fe composition for the two GC RSGs to
Solar abundances, we use the latest Solar Fe composition from the 3-D
Solar model of \citet[][ see Table~{\ref{tab:results}}]{Asplund05}. We
find Fe abundances with respect to Solar [Fe/H] of $+0.14 \pm 0.11$
for IRS~7, and $+0.10 \pm 0.11$ for VR~5-7, where the uncertainties
are the quadrature sum of those in our measurements and those in the
Solar values. Neither of these values represent a
statistically-significant detection of a departure from Solar
abundances. In order to minimise uncertainties, we consider our
analysis in combination with previous studies whose analysis
techniques were complementary to our own. As the studies by
\citet{Carr00}, \citet{Ramirez00} and \citet{Cunha07} were essentially
all made using similar methodology to one another, we do not consider
these three studies as being mutually independent. Instead, we take
the measurements of \citet{Cunha07} to be the state-of-the-art among
these studies. The average of Cunha et al.'s result for IRS~7 and ours
is (${\bar{A}({\rm Fe})} = 7.63 \pm 0.05$), which agrees well with the
average Central cluster value they found. Similarly, for VR~5-7 the
average is (${\bar{A}({\rm Fe})} = 7.58 \pm 0.04$). With respect to
the Solar values of \citet{Asplund05}, this gives [Fe/H]$_{IRS~7} =
0.18 \pm 0.07$ and [Fe/H]$_{VR~5-7} = 0.13 \pm 0.06$.

From these measurements, it appears that there is evidence for slight
Fe enhancement within the massive evolved stars at the GC. The
uncertainties we state above are the simple root-mean-square of the
abundances derived by \citet{Cunha07} and ourselves. It is likely
that, at the uncertainty level of a few percent, we are dominated by
systematic uncertainties which in some way afflict both studies, such
as departures from LTE in the stars' atmospheres. Thus, we conclude
that the two stars studied have surface Fe abundances which are
30-50\% above that of Solar, though at this level we are hitting the
precision limits of stellar abundance analyses.

\begin{figure}[t]
  \centering
  \includegraphics[width=8cm,bb=10 10 546 453]{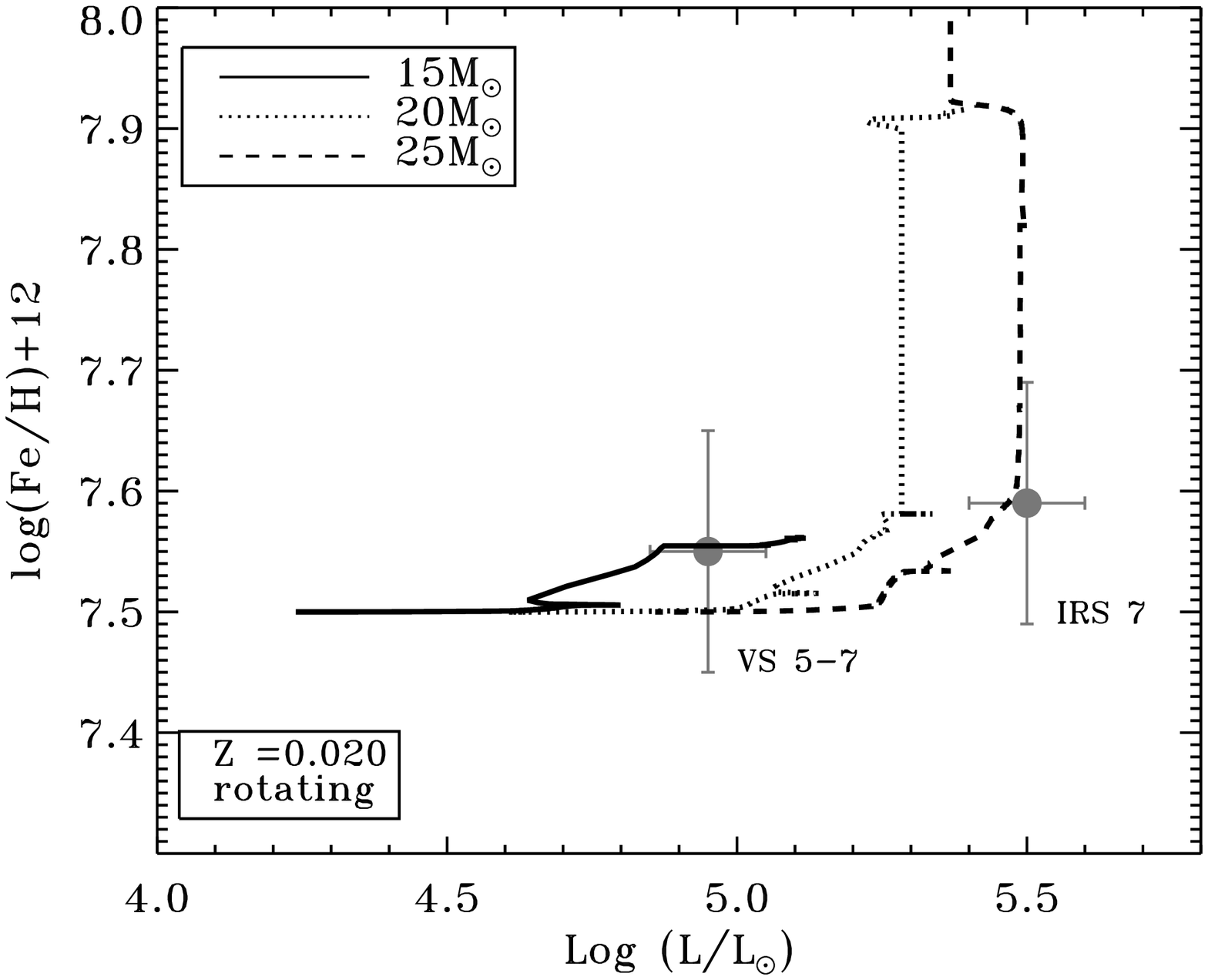}
  \caption{Surface hydrogen depletion of massive stars as a function
  of their evolution, illustrating that the measured Fe/H abundance
  for evolved stars can be larger than that of their initial
  metallicity. The overplotted mass-tracks are from \citet{Hirschi04}. }
  \label{fig:feh}
\end{figure}

However, when interpreting these results one must consider the evolved
nature of the objects studied. As it is likely that the atmospheres of
the stars suffer from H depletion due to core H burning and deep
convective mixing, the {\it initial} Fe abundances of the stars may
well have been entirely consistent with the Solar level. This is
illustrated in Fig.\ \ref{fig:feh}, in a plot of Fe/H against stellar
luminosity for the rotating Geneva stellar evolutionary models
presented in \citet{Hirschi04}. The plot shows how the Fe/H abundance
ratio increases as stars evolve off the main-sequence and the products
of nuclear burning begin to become apparent at the surface. On
average, the surface Fe/H ratio is $\sim$0.1dex lower during the RSG
phase than on the main-sequence.

From comparison with our Fe measurements for the two GC RSGs, it
appears that while both stars currently show marginal evidence for Fe
enhancement, their initial Fe/H composition was likely much closer to
Solar. Similarly, the previous studies of stars in the GC by e.g.\
\citet{Najarro08} and \citet{Cunha07} are consistent with Solar
initial Fe abundances if one considers the evolved nature of the
objects studied.

Finally, we note that a depletion in H due to stellar nucleosynthesis
implies a corresponding increase in He in the stellar envelope. This
has no impact on our analysis method, as the opacity in RSG
atmospheres is fully dominated by H$^{-}$, even if H is marginally
depleted.

\subsection{$\alpha$-elements}
In this study we have determined the abundances of several
$\alpha$-elements: Ca, Si, Mg, O, Ti and C. The interpretation of the
abundances of C and O is not as straight-forward as with the other
$\alpha$ elements, as they are altered by stellar nucleosynthesis. For
this reason we will discuss O separately at the end of this section,
and C in Sect.\ \ref{sec:carb}. For the rest of this study we define
$\alpha \equiv $(Ca, Si, Mg, Ti).

Literature $\alpha$-element studies of these two stars are lacking;
the only previous study being the derivation of Ca abundances by
\citet{Cunha07}. From Table~\ref{tab:results}, we find excellent
agreement with their Ca abundance for VR~5-7. For IRS~7, we find
discrepancies between our two measurements of $\Delta A({\rm Ca}) =
0.4 \pm 0.2$. However, we note that \citet{Cunha07} find an average
$A$(Ca) for the Central cluster of $\bar{A}({\rm Ca}) = 6.57 \pm
0.14$, consistent with our measurement for IRS~7 (${A}({\rm Ca})_{\rm
IRS~7} = 6.46 \pm 0.16$). If the Ca abundances in the Central cluster
are homogeneous, it may be that, in Cunha et al.'s sample, IRS~7
represents a random statistical outlier at the 2$\sigma$ level --
entirely plausible in a sample size of 10. We consider our $A$(Ca)
measurements to be in good agreement with those of Cunha et al.

\citet{Najarro08} studied the $\alpha$-elements Si and Mg in the
spectra of the two LBVs in the Quintuplet cluster, finding marginal
evidence for enhancement with respect to Solar, but with large
uncertainties of 0.2-0.25\,dex. Given these uncertainties, our
measurements of Si and Mg in the spectrum of VR~5-7 are consistent
with those determined by Najarro et al.\ for stars in the same
cluster. 

Overall, on average we find that the $\alpha$-elements studied here
are overabundant with respect to Solar by $\sim0.15$dex. This excess
is clearly within the uncertainties associated with each
measurement. If, for the moment, we assume that each element
represents an independent measurement, we can combine the abundances
for all four $\alpha$-elements studied here to determine the average
$\alpha$-abundance for each star. We find average abundances for each
star of [$\alpha$/H]$_{\rm IRS~7} = +0.18 \pm 0.07$; and
[$\alpha$/H]$_{\rm VR~5-7} = +0.15 \pm 0.05$, where the quoted errors
are the rms scatter of the individual $\alpha$-element
abundances. However, as was described in the previous section, these
abundance ratios may be tainted by $\sim$0.1dex of H self-depletion
due to the stars' evolution. Thus, we consider the {\it initial}
$\alpha$/H abundances of the two stars to be in excellent agreement
with the Solar value.

The parameter of relevence to the study of the GC's star-formation
history is the ratio $\alpha$/Fe: as stated in the introduction,
$\alpha$-elements are enriched by CCSNe, while Fe is enriched by
Type-Ia SN. Hence, a short burst of massive star formation should
produce an over-abundance of $\alpha$-elements with respect to Fe;
while a long star-forming epoch over $\ga$Gyr eventually pulls the
$\alpha$/Fe ratio back down. From the average Fe/H and $\alpha$/H for
each star in this study, we find [$\alpha$/Fe]$_{\rm IRS~7} = -0.04
\pm 0.13$; and [$\alpha$/Fe]$_{\rm VR~5-7} = +0.01 \pm 0.12$. We find
no evidence for $\alpha$-enrichment with respect to iron in these
stars.

For oxygen, comparisons of relative abundances between objects are
hindered by confusion over the Solar O abundance used as a reference
point. In the 1-D Solar model, the O fraction was found to be
$A$(O)$_{\odot} = 8.83 \pm 0.05$, a value which stayed constant to
within the errors over several revisions
\citep{A-G89,G-N93,G-S98}. However, in the 3-D Solar model, the O
fraction has been revised downwards to $A$(O)$_{\odot} = 8.66 \pm
0.05$ \citep{Asplund05}. For the rest of this discussion, we use the
Asplund et al.\ value.

For IRS~7, we find O/H to be depleted by 0.55$\pm$0.14\,dex with
respect to Solar, in good agreement with the study by \citet{Carr00}
of the same star. However, this is much lower ($\sim$0.9dex) than the
average O abundance of the star's host cluster: \citet{Cunha07} found
the Central cluster to have an average super-Solar O/H ratio of
0.38$\pm$0.20\,dex. Interestingly, we find the O abundance of VR~5-7
in the neighbouring Quintuplet cluster to be similar to the average of
Central cluster ($A({\rm O})_{\rm VR~5-7} = +0.43 \pm 0.12$). The
anomolous O abundance of IRS~7 with respect to the rest of the GC RSGs
could be explained as global super-Solar O/H of RSGs in the GC, with
severe O depletion at the surface of IRS~7 due its advanced
evolutionary state -- this stars' evolution is discussed in more
detail in the next section. We find O abundances relative to Fe for
the two objects of [O/Fe]$_{\rm IRS~7} = -0.69 \pm 0.18$; and
[O/Fe]$_{\rm VR~5-7} = +0.33 \pm 0.16$. 

In summary, we find that the $\alpha$ abundances in the GC are
consistent with Solar, both as a fraction of H and Fe. We suggest that
there may be a slight enhancement of O with respect to H, though we
treat this result with caution due to the recent controversy over the
Solar O abundance.

\begin{figure}[t]
  \centering
  \includegraphics[width=8cm,bb=10 10 546 453]{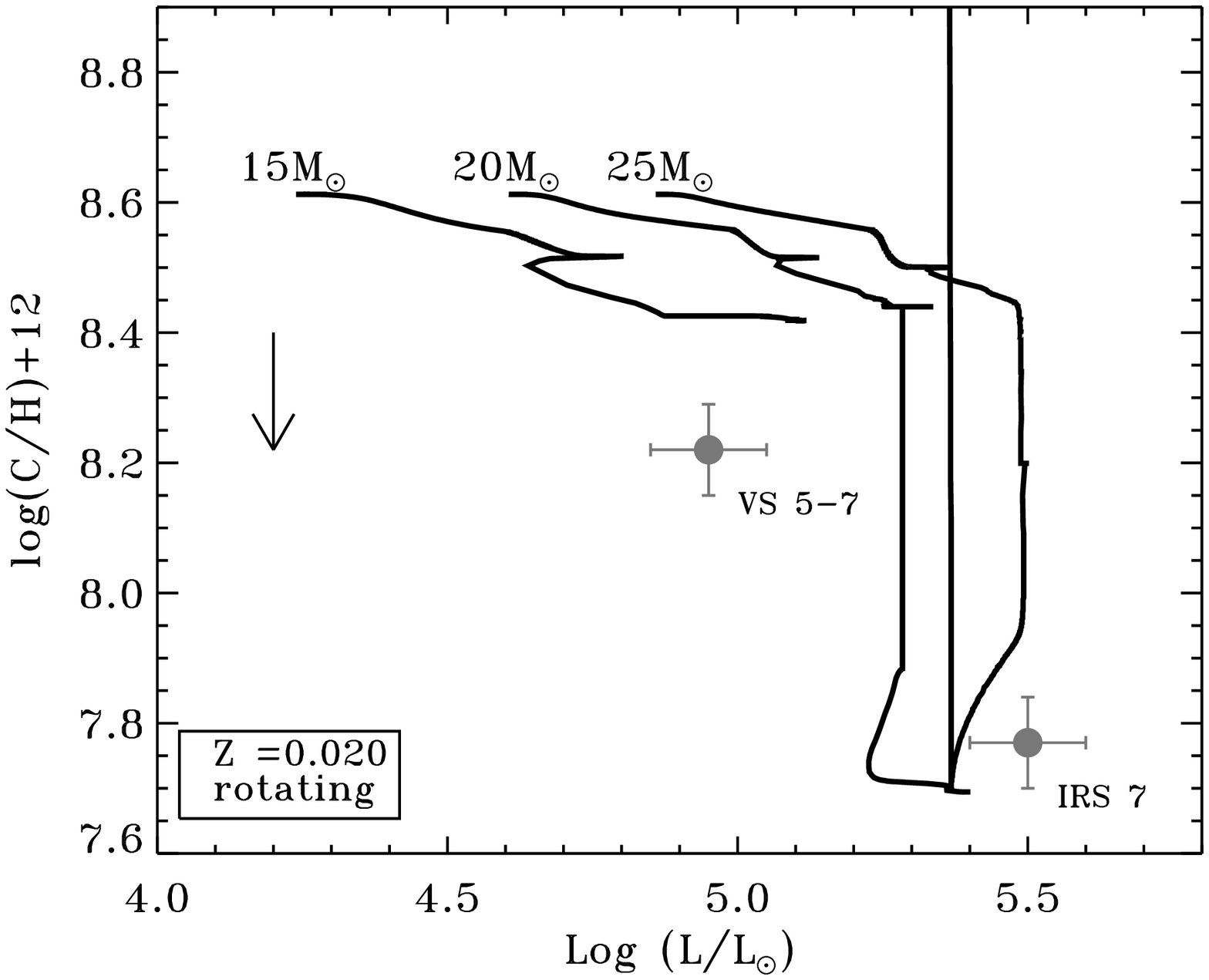}
  \includegraphics[width=8cm,bb=10 10 546 453]{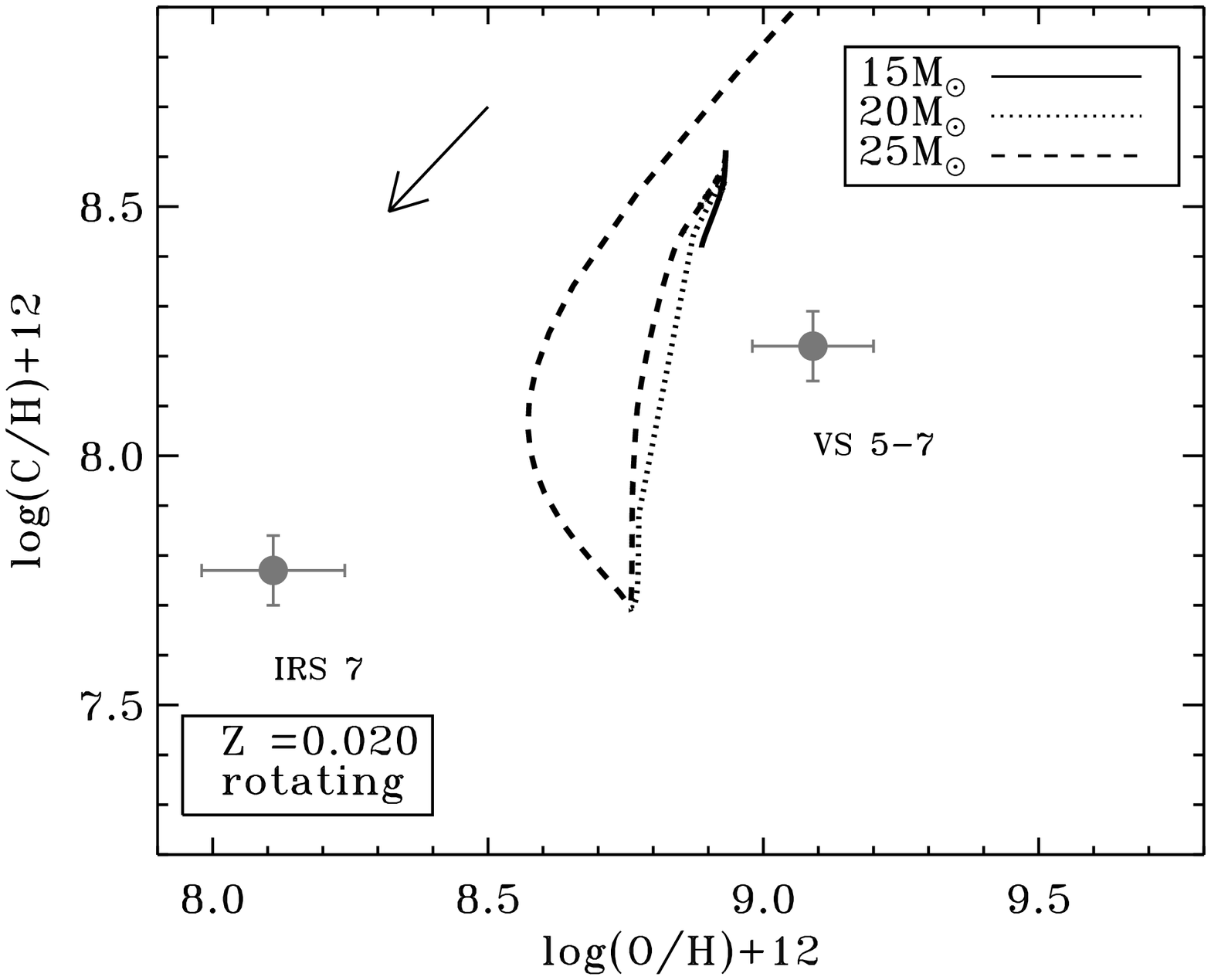}
  \caption{ {\it Left}: The surface carbon abundances of the two stars
  in this study, in comparison to the evolutionary models of
  \citet{Mey-Mae00}. We plot the mass tracks for the three initial
  masses shown, using the Solar metallicity models with initial
  rotational velocities of 300\kms. The arrow denotes the magnitude of
  the change in the updated Solar carbon abundance with respect to
  those used by Meynet \& Maeder in constructing the mass-tracks.
  {\it Right}: Surface C abundances versus surface O abundances. The
  overplotted mass-tracks are from the same evolutionary models as the
  left-hand panel. The arrow again denotes the change in the measured
  Solar composition with the updated Solar model.}
  \label{fig:cl}
\end{figure}

\subsection{Carbon} \label{sec:carb}
As with oxygen, the Solar carbon abundance has also been revised
downwards in the 3-D Solar model, from $A$(C)$_{\odot} = 8.57 \pm
0.05$ \citep{A-G89,G-N93,G-S98} to $A$(C)$_{\odot} = 8.39 \pm 0.05$
\citep{Asplund05}. We use the Asplund et al.\ value in this
discussion. 

Regardless of the adopted value of $A$(C)$_{\odot}$, we find strong
evidence for C depletion in IRS~7. For this object we derive $A$(C)$ =
7.77 \pm 0.07$, in good agreement with the previous study of by
\citet{Carr00}. This represents a carbon abundance relative to Solar
of {[C/H] = -0.62 $\pm$ 0.09}. This is in contrast to VR~5-7, which
is only marginally depleted in C compared to the Solar value. 

In Fig.\ \ref{fig:cl} we plot $A$(C) for each star studied against
their luminosities, which we take from \citet[][]{Cunha07}. To
illustrate the level of C-depletion as a function of the stars'
evolution, we overplot the theoretical mass-tracks of
\citet{Hirschi04}, calculated at Solar metallicity and which
incorporate initial rotational velocities of 300\kms. We note that
these tracks were created using the relative abundances of
\citet{A-G89} and \citet{G-N93}, and therefore use the higher initial
C fractions described above. Thus, all other things being
equal\footnote{We make the assumption that changes of a few $\times$
0.1dex in the relative abundances of C and O will not affect, for
example, reaction rates or envelope opacties such that the
evolutionary path of the star would be significantly altered from
those plotted in Fig.\ \ref{fig:cl}, other than the linear
displacement in abundance-space.}, the tracks should be displaced by
-0.18dex in $A$(C) if the latest relative C abundance is used. This is
indicated by the arrow in the plot panel.

From the figure, we see that VR~5-7 is in good agreement with the
15\msun\ track, even if the track is moved downwards to reflect the
updated C fraction. However, the heavy depletion of C in IRS~7 is not
reproduced by the mass-tracks, even when the lower value for
$A$(C)$_{\odot}$ is taken into account. The sharp downturn in $A$(C)
of the 25\msun\ track corresponds to the removal of the H-rich
envelope and the evolution of the star {\it away} from the RSG stage
towards a Wolf-Rayet phase. On the 25\msun\ mass-track, the minimum
$A$(C) reached {\it during the RSG phase} is similar to the minimum
$A$(C) of the 15\msun\ and 20\msun\ tracks, at $A$(C)$\approx$8.45.

We explore the subject of the stars' evolutionary state further in the
right-hand panel of Fig.\ \ref{fig:cl}, where we plot the carbon
fraction against the oxygen fraction for the two stars. The location
in the panel synonymous with the RSG phase corresponds to the
lower-left end of the 15\msun\ and 20\msun\ tracks at [$A$(O), $A$(C)]
$\approx (8.9, 8.4)$. If, as above, we again make the assumption that
the mass-tracks can simply be displaced to represent the updated Solar
abundances, the star VR~5-7 appears to be enhanced in O with respect
to the evolutionary tracks. IRS~7 on the other hand shows severe
depletion in both C and O, much more than is predicted by the
evolutionary models for an RSG with Solar initial abundances. Indeed,
the position of IRS~7 in $A$(C)-$A$(O) space is more consistent with
the $A$(C) minimum of the 25\msun\ track. In the model, this
corresponds to a nitrogen-rich Wolf-Rayet (WN) phase, where the
products of CNO burning are more readily seen due to the removal of
the outer envelope by the stellar wind. Our results suggest that the
envelope of IRS~7 has experienced a greater level of pollution by
CNO-processed material than is predicted by the standard set of
rotating evolutionary models. This may have been caused by
particularly deep convection zones in the star's atmosphere, or by
enhanced rotational mixing due to a high initial rotation
velocity. Alternatively, a high mass-loss phase could have peeled away
the outer layers, revealing more chemically-processed material, but
without removing enough H to cause the drop the envelope opacity
required for the star to evolve back to the blue.

\subsection{Summary and discussion: chemical abundances in the
  Galactic Centre} \label{sec:ZGC} 
The results of our abundance analyses of two RSGs in the GC can be
summarized as follows:

\begin{itemize}
\item The observed Fe/H ratios of the two stars studied are marginally
  super-Solar, in agreement with previous studies. 

\item For the $\alpha$-elements Ca, Si, Mg and Ti together, we find
  enhancement with respect to H by 40-60\%, albeit at the 2$\sigma$
  confidence level. The $\alpha$/Fe ratios are Solar to within the
  errors ($\pm$25\%), a result which is again consistent with previous
  studies.

\item Weak evidence for O-enrichment is found in VR~5-7, while IRS~7
  is heavily O-depleted - most likely a result of its advanced
  evolutionary state. 

\item We find significant C depletion in the atmospheres of both
  stars. In the case of VR~5-7, this is consistent with the
  predictions of a 15-20\msun\ RSG in the rotating Geneva evolutionary
  models. IRS~7 however has much greater C depletion. In combination
  with this star's low O abundance, this suggests that the products of
  severe CNO processing are been seen at the star's surface. 
\end{itemize}

As summarized in Sect.\ \ref{sec:intro}, several previous studies of
objects in the GC have argued for super-Solar metal abundances in the
GC \citep{Carr00,Ramirez00,Cunha07,Najarro08}. We emphasize here that,
where available, we measure elemental abundances $X/H$ for Fe and
several $\alpha$-elements that are consistent with these
studies. However, in the interpretation of their results these
previous studies did not take into account the self-depletion of H at
the surfaces of the stars due to nuclear burning and internal
mixing. We have shown, with the aid of evolutionary models, that this
level of self-depletion RSGs may be as large as 0.1dex, and is
comparable to the observed level metal enhancement with respect to
H. Hence, the super-Solar $X/H$ ratios may simply be an artifact of
H-depletion rather than being due to enriched metal abundances, and
the {\it initial} $X/H$ abundance ratios of the stars were consistent
with Solar levels. When taking this effect into account, we interpret
the results differently to previous works and conclude that the
metallic elemental abundances in the GC are Solar, at least to within
the precision to which we are currently able to measure them. In
addition, we find that the ratio $\alpha$/Fe is also Solar to within
the errors.

Thus, with a Solar $\alpha$/Fe ratio in the GC it is not necessary to
invoke a peculiar star-forming history, such as a significantly
flattened IMF. Nor do we need to reconcile a super-Solar $\alpha$/Fe
ratio with star-formation which has been continuous for the last
$\sim$Gyr \citep[see Sect.\ \ref{sec:intro}, and ][]{Figer04}. We
again emphasize that our results of Solar $\alpha$/Fe and Solar Fe/H
to within the uncertainties {\it are in agreement with previous
  studies} when H-depletion due to nuclear processing is taken into
account.

So, how do these abundance patterns fit in with the rest of the
Galaxy? A Solar Fe abundance in the GC is certainly inconsistent with
a simple extrapolation of the disk's Fe gradient to lower
Galacto-centric distances. \citet{Luck06} summarize the recent results
of their group and show a clear trend of increasing Fe/H towards the
GC from a large sample of Cepheids. They derive a gradient of
-0.068dex\,kpc$^{-1}$, which when extrapolated all the way to the GC
implies $[{\rm Fe/H}] \approx +0.6$. Our results, as well as those of
other recent studies, show that the Fe/H ratio in the GC is more
consistent with that of the inner disk, at $R_{\rm GC} \sim 5$kpc (see
below).

As noted by \citet{Cunha07}, reconciliation between the disk Fe/H and
that of the GC require that the Fe/H gradient become significantly
flattened inside $R_{\rm GC} \sim 5$kpc. However, there is no reason
to assume that this abundance gradient can simply be extrapolated all
the way to the GC. Shortward of $R_{\rm GC} \sim 5$kpc there is a
distinct lack of recent star-forming activity: there is a dearth of
giant HII regions and massive stars between 0.1kpc$\la R_{\rm GC}
\la5$kpc. Presumably this is related to the presence of the Galactic
Bar, which extends to $R_{\rm GC} \sim 4$kpc \citep{Benjamin05}. This
obvious and sharp transition in star-formation history makes it
unlikely that the abundance gradients in the outer Galaxy are valid
within $R_{\rm GC} \sim 4$kpc. 

We are left with the problem of explaining the abundance patterns in
the GC, with respect to those in the rest of the Galaxy. In external
barred-spiral galaxies, it is commonly accepted that the effect of the
bar is to transport gas from the inner disk to the galaxy's centre --
see for example the observational studies of M100 \citep{Allard06} and
NGC1365 \citep{ZS08}, as well as numerical work
\citep[e.g.][]{Athanassoula92}. Material infalling onto the galactic
centre becomes shocked, and a nuclear starburst is triggered. This
starburst can continue for as long as the centre is supplied with
material. For an extensive review of the secular evolution of barred
spiral galaxies, the reader is directed to \citet{K-K04}.

If the star-formation in the centre of our own Galaxy is being fuelled
by matter transported down the Bar, one would expect to find similar
abundance patterns at the GC as at the edges of the Bar. That is, the
abundances we find in the thin disk at Galacto-centric distances of
$R_{\rm GC} \sim 4$kpc. Abundance studies at these radii are few. The
cepheid data from \citet{Luck06} becomes scarce at $R_{\rm GC} \la
6$kpc with only one data-point at $R_{\rm GC} < 5$kpc. From their
inner-most data they find [Fe/H]$\sim$0.2-0.3dex and
[Ca/Fe]$\sim$0.0. \citet{Smartt01}, \citet{D-C04} and \citet{Munn04}
have observed a total of 6 OB stars within $R_{\rm GC} < 5$kpc,
finding weak evidence for a super-Solar O/H ratio, but with a large
spread. Analysis of the Scutum Red Supergiant clusters, located at
$R_{\rm GC} \approx 3.5-4$\, kpc, has revealed abundances which are
$\approx$0.2dex {\it below} Solar (Davies et al., in prep). In their
discussion of this result in the context of other abundance
measurements in the GC, Davies et al.\ suggest that the abundances at
$R_{\rm GC}$ from 3 to 5\,kpc are approximately Solar, with Solar
$\alpha$/Fe, but with a spread of $\pm$0.25dex. The abundance gradient
between the inner disk and our location in the Galaxy must then be
very shallow, as is typical for galaxies with strong bars
\citep{Zaritsky94}. \citet{Andrievsky02} suggested that the gradient
is extremely flat in the Solar neighbourhood
($\approx$-0.02dex\,kpc$^{-1}$), and steepens at $R_{\rm GC} \la
6$kpc. However, we will show in a forthcoming paper (Davies et al., in
prep) that this steepening may be an artifact of their spatial
sampling, that large azimuthal variations may be present within
$R_{\rm GC} \la 6$kpc, and that this behaviour is seen in the inner
disks of other barred spirals.

The abundances found at $R_{\rm GC} \sim 4$kpc are then similar with
those we find at the GC; namely Fe/H, $\alpha$/Fe and O/Fe which are
all approximately Solar to within the uncertainties of
$\pm$20-30\%. Hence, from a chemistry point-of-view it is entirely
plausible that the star-formation in the GC is fuelled by material
channelled from the inner-disk by the Bar, in a manner similar to that
seen in external barred spirals. Further, the rate at which material
flows down the Bar seems to be consistent with the global
star-formation rate in the GC. \citet{M-S96} use order-of-magnitude
dynamical arguments to estimate the rate of material flowing from the
Inner Lindblad Resonance (ILR) to be $\dot{M}_{\rm ILR} \sim
0.1-1$\msunyr; while \citet{Gusten89} estimate the GC star-formation
rate to be 0.3-0.6\msunyr from the inferred number of Lyman
photons. Hence, the rate of inflowing material appear to be sufficient
to power the GC star-formation, even when one takes into account the
outward-flowing Galactic wind \citep[$\dot{M}_{\rm GW} \la
0.1$\msunyr][]{M-S96}.

Clearly, there are large uncertainties on these numbers, and we cannot
discard the contribution of the winds from Bulge stars. Abundance
studies of these objects have found similar average values of [Fe/H]
and [O/Fe], albeit with a large spread in [Fe/H] of $\pm$0.4dex
\citep{R-O05,C-S06,Lecureur07,Fulbright07}. The total mass-loss rate
from these objects with $R_{\rm GC} \sim 2$kpc has been estimated to
be $\sim$0.1\msunyr, though it is unknown how much of this material
finds its way to the GC. If one wishes to add a further layer of
complexity to the problem, it is possible that metal-rich gas from
recent generations of massive stars mixes with metal-poor gas from the
bulge stars' winds to contrive current abundance levels which appear
approximately Solar \citep[see also][]{vL03}.

In summary, from our abundance analysis of two RSGs in the GC,
combined with the results of other studies, we find that both the Fe
abundance and the $\alpha$/Fe ratio are consistent with Solar values,
and those of the inner Galactic disk at $R_{\rm GC} \sim 4-5$kpc. When
combined with mass-budget arguments, these results suggest that the
recent star-formation in the GC was fuelled by material channelled
along the Galactic Bar from the inner disk, though we cannot rule out
a contribution from the winds of Bulge giants. Thus, there is no need
to invoke of a top-heavy IMF to produce the relative chemical
abundances. However, if a top-heavy IMF {\it does} exist in the GC,
abundance levels in the GC may be held at Solar by a contribution of
metal-poor gas from the winds of bulge giants.


\acknowledgments Acknowledgements: we would like to thank Katia Cunha
and Rob Kennicutt for useful discussions, and Raphael Hirschi for
providing the Geneva mass-tracks at advanced evolutionary stages. We
would also like to thank the referee for a careful reading of the
manuscript and suggestions which helped improve the paper. The
material in this work is supported by NASA under award NNG~05-GC37G,
through the Long-Term Space Astrophysics program. This research was
performed in the Rochester Imaging Detector Laboratory with support
from a NYSTAR Faculty Development Program grant. The data presented
here were obtained at the W.\ M.\ Keck Observatory, which is operated
as a scientific partnership among the California Institute of
Technology, the University of California, and the National Aeronautics
and Space Administration. The Observatory was made possible by the
generous financial support of the W.\ M.\ Keck Foundation. This
research has made use of the IDL software package, and the GSFC IDL
library.

\bibliographystyle{/fat/Data/bibtex/apj}
\bibliography{/fat/Data/bibtex/biblio}

\end{document}